\documentstyle[12pt]{article}

\begin{document}

\begin{titlepage}
\begin{flushright}
YCTP-N11-95\\ June 1995
\end{flushright}
\vspace{1cm}

\begin{center}
{\LARGE Universal  Predictions for Statistical\\
\vspace{2mm}
 Nuclear Correlations}
\end{center} \vspace{.5cm}
\begin{center}
{\large Dimitri Kusnezov\footnote{E--mail: dimitri@nst.physics.yale.edu }
        and David Mitchell}
\\ \mbox{} \\
{\it Center for Theoretical Physics, Sloan Physics Laboratory,} \\
    {\it Yale University, New Haven, CT 06520-8120 USA}

\end{center}

\vskip 1.5 cm
\begin{abstract}
We explore the behavior of collective nuclear excitations under a
multi-parameter deformation of the Hamiltonian. The Hamiltonian matrix
elements have the form $P(|H_{ij}|)\propto 1/\sqrt{|H_{ij}|}\exp(-|H_{ij}|/V)$,
with a parametric correlation of the type $\log \langle H(x)H(y)\rangle\propto
-|x-y|$. The studies are done in both the regular and chaotic regimes of the
Hamiltonian. Model independent predictions for a wide variety of correlation
functions and distributions which depend on wavefunctions and energies are
found from parametric random matrix theory and are compared to the nuclear
excitations. We find that our universal predictions are observed in the nuclear
states. Being a multi-parameter theory, we consider general paths in parameter
space and find that universality can be effected by the topology of the
parameter space. Specifically, Berry's phase can modify short distance
correlations, breaking certain universal predictions.
\end{abstract}
\vspace{1cm}
\noindent


%05.45.+b

\vspace*{\fill}\pagebreak
\end{titlepage}
%%%%%%%%%%%%%%%%%%%%%%%%%%%%%%%%%%%%%%%%%%%%%%%%%%%%%%%%%%%%%%%%%%%%%%%%%%%%%%%
\setcounter{page}{2}
%%%%%%%%%%%%%%%%%%%%%%%%%%%%%%%%%%%%%%%%%%%%%%%%%%%%%%%%%%%%%%%%%%%%%%%%%%

\section{Introduction}

The statistics of nuclear excitations has been explored from the shell model to
collective models, with studies ranging from the relation of observed quantum
fluctuations to  those in random matrix models, to the connection with chaos
using classical limits of the Hamiltonian\cite{general}-\cite{shriner}. The
agreement of various spectral properties with random matrix predictions has
shown that certain simplifying assumptions can be made concerning fluctuations
in nuclei. Once random matrix  theory can be justified, certain results follow
immediately.  These studies of chaos in nuclei stem from attempts to extract a
simplified behavior from the complexity of nuclear excitations.  In this
respect, random matrix theory has provided invaluable assistance in developing
simple methods to compute complex behaviors.
In the past, aside from the studies of constant random matrices and
the relation to chaos, these models have been given a parameter dependence to
model correlations in various nuclear systems, from
heavy ion collisions\cite{weid}, high spin physics\cite{aberg}
to large amplitude collective motion\cite{aurel}.  Recently it has been shown
that Hamiltonians which have a parametric dependence can exhibit universal
behavior\cite{SA}, that is, there can exist model independent quantities in a
given theory, providing the Hamiltonian has  certain random matrix properties.
In this article we study a wide class of observables and develop universal
predictions. We further show that parametric deformations of nuclear
Hamiltonians can be readily modeled by a simple translationally invariant
parametric random matrix theory, even though the Hamiltonian does not apriori
look like a random matrix. We further justify the use of parametric random
matrix theory for collective nuclear excitations.

\section{Collective Nuclear States}

We have chosen to model collective nuclear excitations in the framework of the
Interacting Boson Model (IBM)\cite{franco} for several reasons. One of our main
objectives is to explore and categorize types of model independent  predictions
that exist in parametric quantum theories which exhibit  classical chaos. The
IBM is ideally suited for this since the classical limit has been extensively
studied in recent years using coherent states \cite{Joe}, and the complete
chaotic behavior is now known for every value of the parameters\cite{Niall}.
Hence we can easily choose parametric variations in regions of strong or weak
chaos, or in regular regimes of the parameter space. An additional advantage is
that we can solve the  quantum problem exactly. One might argue that collective
states form only a subset of the real spectrum as the excitation energy
increases, so that the use of the IBM is not reasonable. This is not crucial,
however, since the IBM provides a solvable theory with known spectral
properties, which can be compared to those of the Gaussian Orthogonal Ensemble
(GOE) throughout its parameter range. Certainly a more realistic description of
the spectrum would embody the same features. For example, when broken pair
states are added to the IBM model space, the spectrum becomes more GOE, as the
interactions in the Hamiltonian become more  complicated\cite{francoa}.  This
is certainly the case as one attempts to construct more realistic Hamiltonians.
And as we are showing how {\sl model independent} quantities emerge,  the model
we use is really not so important. Hence we use a simple form of the IBM
Hamiltonian, known as the consistent-Q form:
\begin{equation}           \label{eq:qham}
  \hat{H}= E_0 + c_1\hat{n}_{d} + c_2 {\bf \hat{Q}}^{\chi}\cdot
    {\bf \hat{Q}}^{\chi} + c_3 {\bf \hat{L}}\cdot {\bf \hat{L}},
\end{equation}
where
\begin{equation}
\hat{n}_{d}=d^\dagger\cdot\tilde d\; , \qquad \hat L_\mu =
\sqrt{10}[d^\dagger\times\tilde d]^{(1)}_\mu\; ,\qquad
\hat{Q}^{\chi}_{\mu}=d_{\mu}^{\dagger}s+s^{\dagger}\tilde{d}_{\mu}+\chi
[d^{\dagger}\times \tilde{d}]^{(2)}_{\mu}.
\end{equation}
The parameters $c_i$ are defined by $c_1=\eta/4$ and $c_2=(1-\eta)/4N_b$, where
$N_b$ is the number of bosons. Since the Hamiltonian is diagonalized in a basis
of fixed angular momentum $L$, the constant $c_3$ does not play any role, and
is hence omitted. Except when stated otherwise, we will use  $N_b=25$, which
will give optimal statistics for the quantities we consider. The resulting
dimensions for $J^\pi=0^+,2^+,4^+,10^+$ states are 65,117,165,211. In this
parameterization, one has the following limits: (a) $\eta=1$ corresponds to
vibrational or $U(5)$ nuclei, (b) $\eta=0$ and $\chi=-\sqrt{7}/2$ corresponds
to rotational or $SU(3)$ nuclei, and (c) $\eta=\chi=0$ describes $\gamma-$soft
or $O(6)$ nuclei.

The interpretation of the Hamiltonian in terms of shape variables $\beta$
and $\gamma$ is possible using coherent states, in the  large $N$ limit of
$H$. The energy surfaces for the Hamiltonian in Eq. (1) is
\cite{Joe}
\begin{equation}
 {\cal E}(\beta,\gamma;\eta,\chi) = \beta^2\frac{4-3\eta}{2} +
 \beta^4(1-\eta)(\frac{\chi^2}{14}-1) + \beta^3\cos
 3\gamma\sqrt{1-\frac{\beta^2}{2}}(1-\eta)\frac{2\chi}{\sqrt{7}}.
\end{equation}
For a particular value of $\eta$ and $\chi$, the energy ${\cal E}$ can be
minimized to determine the quantities $\beta$ and $\gamma$. $\beta$ and
$\gamma$ in turn define a deformed nuclear mean field. This can be made
explicit by re-expressing  the Hamiltonian in terms of excitations in a
deformed mean field using boson condensate techniques\cite{ami}. This
allows the interpretation of correlations of observables at different values of
$\eta$ and $\chi$ in terms of the shape variable $\beta$ and $\gamma$.
Correlations in observables at different values of the  parameters are then
precisely the correlations between properties of the nucleus in the presence of
different mean field configurations. We will consider the behavior of the
properties of the Hamiltonian under very general parametric deformation
$z=z(\eta,\chi)$. For paths which lie entirely within the chaotic regime of the
parameter space, the universal predictions we explore are path independent (up
to effects due to Berry's phase which we explore in Sec. 5); correlations in a
nucleus changing from rotational to vibrational or vibrational to $\gamma-$soft
are the same when properly interpreted.

\subsection{Distributions and Correlations of Nuclear Matrix Elements}

One of the results presented in this article is that parametric nuclear
Hamiltonians can be modelled by correlated, parametric gaussian random
matrices. Recall that a gaussian random matrix has a distribution of matrix
elements of the gaussian form $P(H_{ij})\propto
\exp(-H_{ij}^2/2\gamma(1+\delta_{ij}))$, where  $\gamma$ a constant related to
the level density. To implement random matrix theory does not imply that the
actual nuclear Hamiltonian (1) have gaussian matrix elements. We note that the
distributions of matrix elements of the interacting boson model Hamiltonian are
not gaussian. At any given value of $(\eta,\chi)$, we find the distribution of
matrix elements obeys roughly\cite{flam}
\begin{equation}
  P_{ibm}(|H_{ij}|) \propto \frac{1}{\sqrt{|H_{ij}|}} e^{-|H_{ij}|/V}
\end{equation}
where the strength $V$ depends on whether one is in a chaotic or regular
regime. Typical results are shown in Fig. 1 for both regular (crosses) and
chaotic (boxes) choices of the parameters, together with the behavior (4)
(solid curves). In the chaotic parameter regimes of the model, $V$ is of order
unity, while in the regular regions, it is much smaller. But both regular and
chaotic regimes display the same functional form of the distribution,
suggesting that the functional form is due to the structure of the Hamiltonian,
rather than to the presence of chaos. Similar distribution functions have been
seen in parity non-conservation studies of the compound nucleus\cite{flam}.

Another quantity of interest is the autocorrelation function $F_{ibm}$ computed
from the IBM Hamiltonian:
\begin{equation}
  F_{ibm}(z-z') \equiv \left\langle H(z)H(z')\right\rangle =
  \left\langle\frac{1}{N(N-1)}\sum_{i<j}
                                H_{ij}(z)H_{ij}(z')\right\rangle_{z-z'}.
\end{equation}
The brackets $\langle \cdots\rangle_{z-z'}$ represent the averaging over a
trajectory in parameter space $z(\eta,\chi)$, which remains either in a chaotic
or in a regular region, keeping the difference $z-z'$ fixed.
The results for the short distance behavior of the measured function
$F_{ibm}$ are shown in Fig. 2, and are found to behave generically as
\begin{equation}
  F_{ibm}(z) \sim e^{-\gamma |z|} \sim 1 - \gamma |z| + \cdots
\end{equation}
in both regular and chaotic regions. Again, the measured value of $F_{ibm}$ is
not a good measure of the underlying chaos.  If the averaging in (5) is
restricted to a submatrix $N_1 \leq i,j\leq N_2$, there is no difference in the
function $F_{ibm}$. One observation is that the apparent decorrelation, seen in
the  slow decay of $F_{ibm}$, is misleading. The actual decorrelation is much
more rapid (when model specific dependencies are removed), as we will see
below when we compute properties of wavefunctions and eigenvalues. Hence using
Eq. (5) as actual input into a random matrix theory (e.g. into Eq. (16) below)
is not meaningful.

\subsection{Unfolded Parametric Energies}

To study the statistical fluctuations of the nuclear energy levels, $E_n(z)$,
where $z=z(\eta,\chi)$ is a general path in configuration space, we must
separate out the average behavior of the energies which cannot be described by
random matrix theory. This is done with the staircase function\cite{general}
\begin{equation}
 N(E;z)= {\rm Tr}\theta(E-\hat H)=\sum_n \theta[E-E_n(z)]
\end{equation}
This function is computed along various paths in parameter space. On each
path, 100-200 values of $z$ are taken, and the eigenvalues $\{E_n(z)\}$
determined. From this data, a polynomial fit is made to the staircase
function using:
\begin{equation}
 N(E;z) = \sum_{i=0}^{k}\sum_{j=0}^{6} C_{ij} z^i E^j.
\end{equation}
In the chaotic regions of parameter space, $k=2$ is sufficient, whereas
higher values are needed is less chaotic regions.
Once the coefficients $C_{ij}$ are determined, the {\sl unfolded energies}
are defined by
\begin{equation}
 \widetilde E_n(z) = N(E_n;z)
\end{equation}
which results in a spectrum with a constant average level spacing. The degree
of chaos in the energies can be measured through the Brody distribution of the
level spacings $s= \widetilde E_n(z) -  \widetilde E_{n-1}(z)$\cite{brody}:
\begin{equation}
 P(s) = A s^\omega e^{-\alpha s^{\omega+1}},\qquad\qquad
  A=(1+\omega)\alpha,\quad  \alpha=\left[\Gamma
  \left(\frac{2+\omega}{1+\omega}\right)\right]^{1+\omega}.
\end{equation}
When $\omega=1$, the distribution of level spacings is GOE, while for
$\omega=0$, it is Poisson. In Fig. 3, we show the original (top) and unfolded
(middle) parametric energies of the Hamiltonian for $\eta=0$ and $\chi$ as
shown. The degree of `chaos', measured in terms of the Brody parameter
$\omega$, is shown in the bottom of the figure for this particular path.

For purposes of contrast, the parametric levels (not unfolded) in two regular
regions are shown in Fig. 4. The right figure corresponds to the (regular)
transition from $U(5)$ to $O(6)$, while the left corresponds to a path through
the `valley of regularity' recently studied\cite{Niall} (the two kinks in the
parametric energies are artificial and only reflect the fact that the parameter
path took a turn). In the figure, the average Brody parameters
$\overline{\omega}$ are 0.25 and 0.23, respectively, with fluctuations up to
$\omega=1.$ In the following studies, we will consider regular and chaotic
regions of the parameter space. As chaos is a classical notion and we study
quantum statistics here, let us be precise. The chaotic regions are general
paths in parameter space $z(\eta,\chi)$ which stay in the areas of chaos in the
classical limit of the Hamiltonian, as discussed in Ref. \cite{Niall}. The
chaotic paths studied here are largely:
\begin{equation}
 z(\eta,\chi) = \left\{\begin{array}{l}
   \eta=0,\quad \chi\in [-0.9,-0.4]\\
   \eta=0.1,\quad \chi\in [-0.9,-0.4]\\
   \chi=-0.66,\quad \eta\in [0.0,0.5]
\end{array}\right.
\end{equation}
Similarly, the regular regions we study are:
\begin{equation}\label{eq:pathr}
 z(\eta,\chi) = \left\{\begin{array}{ll}
   \chi=-\frac{\sqrt{7}}{4}\eta\qquad \eta\in[0,1] & O(6) - U(5)\\
   \chi=-0.661,\quad \eta\in [0.5,1.0] & \\
\end{array}\right.
\end{equation}
Also included in the regular region is the path through the valley of
regularity from $SU(3)$ to $U(5)$, as shown in the left half of Fig. 4. The
first path in Eq. (\ref{eq:pathr}) corresponds to the right half of Fig. 4, is
not the direct path between $U(5)$ and $O(6)$ which would be $\chi=0$, $\eta\in
[0,1]$, and is entirely integrable. Rather we have chosen one that passes
through a weakly chaotic regime. Hence, results in the following sections
referred to as `regular' are  quantities which are averaged in the highest
$\omega$ regions of these regular areas, such as $z\in[0.5,0.8]$ in Fig. 4.

\section{Correlated Random Matrix Ensembles}

In order to see what types of model independent quantities emerge from the
IBM, we must construct a random matrix model which has an equivalent
parametric dependence. There is no unique method to realize such an ensemble,
for instance one might take:
\begin{eqnarray}
  H'(z) & = & H_1 + z H_2\\\label{eq:simple}
  H''(z) & = & \sin z H_1 + \cos z H_2\\\label{eq:cosine}
  H'''(z) & = & \int dy f(z-y) V(y),\qquad \overline{V(z)V(y)}=\delta(z-y)
  .\label{eq:stoch}
\end{eqnarray}
Here $H_1$, $H_2$ are constant $N\times N$ gaussian random matrices and
$V_{ij}(y)$ is
gaussian white noise for each $i,j$ and $y$. Each of these is a viable random
matrix theory, however the stochastic integral of Eq. (\ref{eq:stoch}),
introduced by Wilkinson\cite{wilk},  provides a more general framework and
includes a broader class of processes\cite{caio}. One additional difference
between $H'$ and $H''$, $H'''$ is that the former is not a translationally
invariant theory. While translational invariance is not important to the
results we derive here, its presence simplifies our constructions.
We will only focus on constructions of the type (14)-(15) here.

The gaussian random matrix Hamiltonians are characterized by their first and
second cumulants:
\begin{eqnarray} \label{eq:aver}
   \overline {H_{ij} (z) } & = & 0 \nonumber \\
   \overline{ H_{ij} (z) H_{kl} (z^\prime) }  & = & {a^2 \over {2\nu}}
F(z-z^\prime) g^{(\nu)}_{ij,kl} \;,
\end{eqnarray}
where $g^{(\nu =1)}_{ij,kl} = \delta_{ik} \delta_{jl} + \delta_{il}
\delta_{jk}$, $g^{(\nu =2)}_{ij,kl} = 2 \delta_{il} \delta_{jk}$, and $a$
determines the average level spacing $\Delta$ through the relation $a/\Delta =
\sqrt{2N}/ \pi$.  Here $\nu=1$ (GOE) corresponds to real-symmetric matrices, or
equivalently, to a system with time reversal symmetry, and $\nu=2$ (GUE) to
complex hermitian matrices, or broken time reversal symmetry. From the
definitions of $H(z)$, it is clear that $H(z)$ is GOE $(\nu=1)$ or GUE
$(\nu=2)$ for any $z$.

In contrast to previous studies of chaos in nuclei, which deal with constant
random matrices, we can introduce a measure for the parametric ensemble rather
easily\cite{caio}:
\begin{equation}
   P\left[ H(z)\right] \propto
   \exp \left\{-{\nu \over {2a^2}}\int dz dz^\prime
    {\rm Tr} \left[ H(z) K(z-z^\prime) H(z^\prime) \right]  \right\} \;,
\end{equation}
where the measure $D\left[ H(z)\right] \equiv \prod_z{dH(z)} $ is a product
over the continuous variable $z$ of the corresponding gaussian ensemble measure
$dH(z)$. Here $K_{ij}(z)$ can be viewed in general as a banded matrix of
bandwidth $\sigma$, connecting states $i$ and $j$ with $|i-j|\leq \sigma$. As
we do not consider banded parametric matrices here, we will take
$K_{ij}(z)=\delta_{ij} K(z)$, resulting in the measure:
\begin{equation}
   P\left[ H(z)\right] \propto
   \exp \left\{-{\nu \over {2a^2}}\int dz dz^\prime K(z-z^\prime)
    {\rm Tr} \left[ H(z) H(z^\prime) \right]  \right\} \;,
\end{equation}
These gaussian integrals are easily done to establish Eq.
(\ref{eq:aver}), providing $F$ is the inverse of $K$,
\begin{equation}
 \int dz^\prime  K(z-z^\prime)F(z^\prime - z^{\prime\prime})
 = \delta(z-z^{\prime\prime}).
\end{equation}
The stochastic integral (\ref{eq:stoch}) provides a direct method for
constructing $H(z)$ with a desired $F(z)$. That is, we can choose $f$
to satisfy
\begin{equation}
  F(z-y) = \int dx f(z-x) f(y-x),
\end{equation}
then $H$ is constructed as in (\ref{eq:stoch}), and  the desired covariance
(\ref{eq:aver}) is
automatically satisfied. It is important to realize that the properties of the
random matrix theory here are distinct from the observed properties of the
Hamiltonian, both in the measured distributions of matrix elements
$P_{ibm}(H_{ij})$ and their autocorrelation $F_{ibm}(z)$. The random matrix
distribution of matrix elements is gaussian, and $F(z)$ is different from
$F_{ibm}$, as we discuss below. In particular, $F(z)$ cannot be exponential as
measured. Model independent results can be obtained from our random matrix
constructions by a proper scaling of parameters. We discuss two approaches here
to this scaling. The first is a general procedure based on Ref. \cite{caio},
while the second is a more heuristic argument based on the Fokker-Planck
equation and the original work of Dyson, and has been pursued in recent
works\cite{dyson,fokker,walk}.

\subsection{Scaling from Anomalous Diffusion}

It was shown recently that universal (model independent) predictions can be
obtained from the above translationally invariant random matrix theories
if one introduces a proper
scaling\cite{caio}. A general approach to do so is to view the parametric
dependence of the energies as a diffusion process.  Consider first the short
distance behavior of the function $F(z)$:
\begin{equation}\label{eq:falpha}
  F(z) \approx 1 - c_\alpha |z|^\alpha + \cdots\qquad .
\end{equation}
{}From perturbation theory, one can see that
\begin{eqnarray}
\label{eq:diff}
  \delta E_n(x)&=& E_n(x')-E_n(x) = \delta H_{nn}
      + \sum_{m\not= n} \frac{|\delta H_{mn}|^2} {E_n-E_m} + ...\\
  \mid\langle\Psi_n (x')|\Psi_n  (x)\rangle\mid^2 &=&
    1-\sum_{m\not= n}\frac{|\delta H_{mn}|^2}{(E_n-E_m)^2} + ... \, .
\end{eqnarray}
By implementing the ensemble averages defined in (16), and following
Dyson\cite{dyson}, one easily finds that:
\begin{equation}\label{eq:ediff}
\overline{(\delta E_n(z))^2} \simeq \frac{4Nc_\alpha}{\nu\pi^2}\delta
z^\alpha  \equiv D_\alpha \delta z^\alpha\quad,
\end{equation}
and similarly
$1-\overline{\mid\langle\Psi_n(z)\mid\Psi_n(z')\rangle\mid^2}\propto\delta
z^\alpha$. One can then view the parametric energy levels $\widetilde{E_n}(z)$,
such as those in Fig. 2 (middle), as evolving diffusively on short distance
scales according to Eq. (\ref{eq:ediff}). This has been recently contrasted
with the anomalous diffusion  process of a particle in a chaotic or disordered
medium, whose position obeys $<R^2(t)> = Dt^\alpha$, which although the physics
is distinct, the formal treatment is similar\cite{caio}.  For our random matrix
model, the parameter $z(\eta,\chi)$ plays the role of time. The diffusion
constant $D_\alpha$ contains both dimensional information ($N$) and model
dependent data $(c_\alpha)$. Hence, by scaling  the parameter $z$ by the
diffusion constant $D$, all model and dimension dependence is removed. This is
done by defining a new scaled parameter
\begin{equation}
 \widetilde{z} = [D_\alpha]^{1/\alpha} z
 = \left(\frac{4Nc_\alpha}{\nu\pi^2}\right)^{1/\alpha} z.
\end{equation}
For the case of $\alpha=2$, we have $D_2=C(0)$, where $C(0)$ is the scaling
introduced in Ref. \cite{SA}. By computing observables in the scaled variable,
one obtains model-independent predictions for desired quantities. Physically,
$\alpha=2$ in Eq. (\ref{eq:falpha}) corresponds to a Hamiltonian with a  smooth
dependence on the parameter $z$, while $\alpha<2$ corresponds to a  theory with
fractal parametric dependence, such as a parameter range taken from a Brownian
trajectory. The computation of $D_\alpha$ is done through the definition
(\ref{eq:ediff}):
\begin{equation}
 D_\alpha = \overline{\frac{(\delta E_n(z))^2}{\delta z^\alpha}},\qquad\qquad
 C(0)\equiv D_2 = \overline{\frac{(\partial E_n(z))^2}{\partial z^2}}.
\end{equation}

In order to model the IBM Hamiltonian using parametric random matrix theory, we
must choose a correlator  $F(z)$ in Eq. (\ref{eq:aver}) with $\alpha=2$ short
distance behavior (see Eq. (\ref{eq:falpha})). Otherwise the parametric
dependence of the random matrix energies $E_n(z)$ would not be smooth. Hence,
if we attempt to incorporate nuclear properties into the random matrix model by
substituting the model specific, computed $F_{ibm}(z)$ in Eq. (6) into Eq.
(\ref{eq:aver}), we would end up with an energy spectrum $E_n(z)$ characterized
by $\alpha=1$, resulting in non-smooth, brownian motion type paths for each
$E_n(z)$. For the IBM, $\alpha=2$ is the proper result.  Details
can be found in Ref. \cite{caio}.

\subsection{Scaling and the Fokker-Planck Equation}

A more heuristic argument can also be made with the Fokker-Planck equation.
Fokker-Planck methods, first introduced by Dyson\cite{dyson}, have been
recently discussed with parametric correlations in
mind\cite{fokker}-\cite{walk}. Consider first Dyson's Brownian motion model
for random matrices:
\begin{equation}\label{4}
   \dot{H}_{ij} = - \gamma H_{ij} + f_{ij}(t) \quad .
\end{equation}
The random force is white noise:
\begin{eqnarray} \label{5}
    \overline{ f_{ij} (t) } & = & 0  \nonumber \\
    \overline{ f_{ij}(t) f^*_{kl} (t') } & = & \Gamma g^{(\nu)}_{ij,kl}
                                     \delta (t - t')  \;.
\end{eqnarray}
and $\gamma$ is a friction coefficient. The equilibrium solution can be found
in the long time limit, by  direct integration:
\begin{equation}\label{eq:integ}
  H_{ij}(t) = \int^t e^{-\gamma|t-\tau|}f_{ij}(\tau)d\tau
\end{equation}
which is same type of stochastic formulation as in Eq. (\ref{eq:stoch}).
It follows then from Eqs. (\ref{4},\ref{5},\ref{eq:integ}) that:
\begin{equation}\label{eq:ave}
  \left\langle H_{ij}(t)H_{kl}(t')\right\rangle
   = \frac{\Gamma}{2\gamma} g^{(\nu)}_{ij,kl} e^{-\gamma |t-t'|}
\end{equation}
This process can be formulated as well in terms of the Fokker-Planck equation
for the distribution $P(H,t)$:
\begin{equation}\label{6}
  \frac{\partial P}{\partial t} = \frac{\partial}{\partial H_{ij}}
 (\gamma H_{ij} P) +
  \frac{1}{2} g^{(\nu)}_{ij,ji}  \Gamma \frac{\partial^2 P}
                              {\partial H_{ij} \partial H_{ij}^*}  \;.
\end{equation}
Since we are interested in a stationary process $P(H,t)=P(H)$ for $H(t)$,
we can choose the initial distribution to be the equilibrium result
$P(H)\propto \exp [- \nu {\rm Tr}H^2/2 a^2 ] $.
The equilibrium solution is a solution of the Fokker-Planck equation providing
the fluctuation dissipation theorem is satisfied:

\begin{equation}
 g^{(\nu)}_{ij,ji}\Gamma/2 \gamma = \overline{\mid H_{ij} \mid^2 } =
 \frac{a^2}{2\nu} g^{(\nu)}_{ij,ji}
\end{equation}
or $a^2/\nu = \Gamma/\gamma$, where the last equality follows from our
original construction in Eq. (\ref{eq:aver}). Following Dyson's argument that
the eigenvalues of $H(t)$ behave as a diffusive coulomb gas, equilibrating (at
the microscopic scale) on a timescale of $t\propto 1/(\gamma N)$, we see that
in  order to obtain $N-$independent correlations, we must have $\gamma\propto
1/N$, or $\gamma=\gamma'/N$. By equating the second cumulants of $H(t)$ for and
Langevin process (\ref{eq:ave}) and for our desired process (\ref{eq:aver}),
and implementing the fluctuation-dissipation theorem, we equate
\begin{equation}\label{eq:compare}
   F(z-z^\prime) \leftrightarrow \exp{ \left( - \frac{\gamma ^\prime}{N}
 \mid t-t^\prime \mid \right) }  \;.
\end{equation}
Expanding to leading order in the large $N$ limit, we associate:
\begin{equation}
  1-c|z|^n\simeq 1-\frac{\gamma'}{N}|t|
\end{equation}
As the Langevin result is $N-$independent at the microscopic scale, $N$
independent results will also result if we define a new quantity
$\hat{z}=(c/N)^{1/n} z$. Up to a factor of order unity ($4/\pi^2\nu$), this is
precisely the same scaling we discussed earlier. Now to leading order,
\begin{equation}\label{12}
F\approx 1 - \nu \frac{\pi^2}{4} \frac{\mid\widetilde{z} -
\widetilde{z}^\prime \mid^n }{N} \;.
\end{equation}
There is now no explicit model dependence $c$, and the $N-$dependence of
correlation functions will be absent due to the explicit dependence on $N$,
which was require to achieve the microscopic equilibrium condition discussed
above.

\section{Observables}

In this section we will see that many properties of parametric Hamiltonians
have well defined model independent structure.
There are two classes of observables we study here, those related to the
energies $E_n(z)$ and those related to the instantaneous eigenfunctions
$\mid\Psi_n(z)\rangle$.
The relation of wavefunctions $\mid\Psi_n(z)\rangle$ at $z$ to those at $z'$ is
given by the transformation matrix:
\begin{equation}
  U_{nm}(z-z') = \langle\Psi_n(z)\mid\Psi_m(z')\rangle
\end{equation}
We will see below that correlation functions that depend on $U_{nm}(z)$ and
$E_n(z)$ have universal predictions which generally agree well with the
results of the IBM in the chaotic regions.

The universal predictions for an $\alpha=2$, GOE system are computed here with
two different covariances. The first is the simple sum of two uncorrelated
GOE matrices
\begin{equation}\label{eq:fone}
 H'(z) = H_1 \cos{z} + H_2 \sin{z},\qquad
       \overline{H'(z)H'(y)}=\cos(z-y)=F'(z-y).
\end{equation}
The correlator is periodic with $F'(z)\approx 1-z^2/2 + \cdots$ ,
defining the scaling $z \rightarrow \widetilde{z}=\sqrt{D_2}z=(\sqrt{2N}/\pi)
 z$ for universal correlations. The second
construction is in terms of a stochastic integral, where one integrates over a
continuous range of uncorrelated random matrices $V(y)$:

\begin{equation}\label{eq:ftwo}
  H''(z) = \int dy e^{-(z-y)^2/2} V(y),\qquad
          \overline{H''(z)H''(y)}=e^{-(z-y)^2/4}=F''(z-y)
\end{equation}
Here $\widetilde{z}=(\sqrt{N}/\pi)z$. We have computed various correlation
functions described below with $N=50-300$. Generally $N=50$ is sufficiently
large.

\subsection{C(z)}

The slope-slope correlation function of the unfolded parametric energies
$\widetilde{E_n}(z)$ is defined as\cite{SA}
\begin{equation}
C(z-z') = \left\langle \frac{\partial \widetilde{E_i}(z)} {\partial z}
  \frac{ \partial \widetilde{E_i}(z')}{\partial z}\right\rangle_{E,z}
\end{equation}
where the averaging is over energy and parameter. In Fig.5, the results for the
IBM in the chaotic regions are shown for $J^\pi=0^+,2^+,4^+,10^+$. The
averages are computed by averaging over the middle third of the spectrum and
over the trajectory $z$. The two solid lines are the results of the random
matrix simulations, Eqs. (\ref{eq:fone}-\ref{eq:ftwo}), for $N=50$ and 300. For
comparison, a computation in the regular region of the IBM is shown indicating
much slower decorrelation. As we will see in all computations here, the typical
distance at which quantities decorrelate is $\widetilde{z}\sim 1$, which
corresponds to the average separation between level crossings when the energies
$\widetilde E_n$ are plotted as a function of the scaled parameter $\widetilde
z$. For regular systems, and the apparent level crossings in Fig. 4, this is
not the case, and  decorrelation happens over a much longer scale. In general
the agreement with the universal functions is quite good. The $0^+$ states have
the poorest agreement, which is in part statistical, having the smallest
dimension.

\subsection{N-Scaling of C(0)}

We have seen that the diffusion constant $C(0)=D_2$ scales linearly with $N$,
the dimension of the space. This scaling can be tested in the IBM by modifying
the boson number $N_b$. For $N_b=10,15,20,25$ we have dimensions of $J=10^+$
states of $N=16,56,121,211$. In Fig. 6 we plot $C(0)$ as a function of $N$, and
see that this scaling is observed. At low $N_b$ (equivalently low $N$), the
results are not as reliable due to statistics becoming increasingly poor, and
the classical phase space becoming increasingly regular\cite{Niall}.

\subsection{Curvature Distribution P(k)}

The distribution of curvatures of the parametric energies $E_n(z)$, or equally
$\widetilde{E_n}(z)$, have a predicted distribution in the chaotic regime,
given by\cite{delande}:
\begin{equation}
 P(k) = \frac{c_\nu}{(1+k^2)^{(\nu/2+1)}},\qquad\qquad k =
 \frac{d^2 \widetilde{E_n}}{dz^2}\frac{1}{\pi\nu C_0}.
\end{equation}
Here $c_\nu$ is the normalization, $\nu=1(2)$ for GOE(GUE), and $k$ is the
scaled curvature of the parametric energy. This function is compared to our
random matrix computation (Eq. (\ref{eq:ftwo})) in Fig.7 (a). In Fig. 7 (b),
the results for the chaotic region of the IBM (solid histogram) are seen to
agree  equally well. A similar calculation done in the regular region shows a
much  more strongly peaked function (dashed, and scaled by 1/5 vertically).
This is  expected since the regular regions have much fewer level crossings and
are hence  much flatter (see Fig. 4).

\subsection{Diagonal Wavefunction Decorrelations: $P_{n}(|U_{nn}(z)|^2)$}

The adiabatic survival probability $|\langle\Psi_n(z)\mid\Psi_n(0)\rangle|^2$
measures how rapidly wavefunctions decorrelate. This was shown to be universal
recently\cite{caio,david}, with a well defined Lorentzian shape:
\begin{equation} \label{eq:psis}
 P_n(|U_{nn}(z)|^2) = \overline{|\langle \Psi_n(z)|\Psi_n(y)\rangle|^2}
    = \left(\frac{1}{1+c|\widetilde{z}-  \widetilde{y}|^\alpha}\right)^\nu
\end{equation}
where $\nu=1(2)$ corresponds to GOE(GUE) eigenstates, $c$ is a constant, and
$\alpha$ is given by the leading order behavior of $F(z)$. For the case at
hand, $\nu=1$ and $\alpha=2$. In Fig. 8, comparisons are shown for selected
spins, and in chaotic (symbols) and regular (dashed) regions. The regular
regions indicate much longer correlations, while the GOE result provides the
most rapid statistical decorrelation of states. The two solid lines are the
random matrix predictions. There is no equivalent universal prediction for
the non-chaotic regimes, and the dashed line is just a representative $P_n$
of the IBM in the regular region.

\subsection{Off-Diagonal Wavefunction Decorrelation: $P_{k}(|U_{nm}(z)|^2)$}

Wilkinson and Walker\cite{walk} have used perturbation theory to derive an
approximate expression for the distribution of squared off-diagonal matrix
elements, $|\langle \Psi_n(z)\mid\Psi_m(0)\rangle|^2$, in the limit of
$|z|\rightarrow\infty$ and $k=|m-n|\gg 1$. They found that
\begin{equation}\label{eq:wilk}
  P_{nm}(z) = \frac{\mu^2z^2}{ (E_n-E_m)^2 + (\pi\rho\mu^2
  z^2)^2},\qquad\mu^2 = \left\langle |\frac{\partial H}{\partial
  z}|^2\right\rangle_{E_n\sim E_m,n\not= m}
\end{equation}
Here, the energies are not averaged over, $\mu^2\sim 1$, and $\rho$ is the
average level density.  In exploring the behavior of these quantities in the
IBM, it is difficult to satisfy the validity conditions for Eq.
(\ref{eq:wilk}), since we largely  study states in the middle portion of the
spectrum, and $m\gg n$ is difficult to satisfy. If we use only two states
separated by $k=|n-m|\gg 1$, and we do not average over energy, the statistics
are very poor. In order to get good statistics, we have examined the equivalent
distribution which is averaged over both coordinate $z$ and energy, with $k$
kept fixed. We define this distribution of off-diagonal matrix elements as
$(k>0)$:
\begin{equation}
  P_k(\widetilde{z}) = \left\langle |U_{nm}(z)|^2\right\rangle
                    = \left\langle |\langle\Psi_n(z+z_o)|
        \Psi_m(z_o)\rangle|^2\right\rangle_{E,z_o,k=|n-m|}
\end{equation}
where the subscript indicates that it is averaged over the trajectory
$z(\eta,\chi)$, as well as over energy, with the separation $|n-m|$ held fixed.
We use the notation $P_{nm}$ for the quantity which is not energy averaged, and
$P_k$ for the energy averaged result.
In Fig. 9, we compare this function computed in random matrix theory (solid)
and in the chaotic regime of the IBM (boxes).  In general the agreement is
quite good. The random matrix theory result was done using $N=50$, and
averaging over the middle third of the spectrum. Hence as $k$ increases, the
statistics get worse. The IBM results are for the $J^\pi=10^+$ states,
with a dimension of 211, so that the statistics is better. In order to contrast
our results with Eq. (\ref{eq:wilk}), we have taken a rescaled form, which is
not entirely justified, as the scaling by $\mu$ is distinct from $D_2=C(0)$.
Nevertheless, we plot in Fig. 9, the following rescaled functions (whose
regimes of validity are indicated):
\begin{eqnarray}
 P_k'(\widetilde{z}) & = & \frac{\widetilde{z}^2}{k^2 + \widetilde{z}^4},\qquad
                              k\gg 1, |z|\gg 1\\
 P_k''(\widetilde{z}) & = & \int_0^\infty dx \cos (kx)
           e^{-\widetilde{z}^2[1-\exp(-|x|)]},\qquad k\gg 1
\end{eqnarray}
The function $P_k'$ (dots in Fig. 9) is an approximation of $P_k''$
(dot-dashed in Fig. 9) in the limit $|z|\gg 1$, which was derived in Ref.
\cite{walk}. As the value of $k$ increases, there is better agreement with
the exact results form the random matrix simulation and the IBM.
We observe that a better overall fit can be found with the function
(dashed line if Fig. 9)
\begin{equation}
 P_k (\widetilde{z}) =
 \frac{\widetilde{z}^2}{k(k-c) + \widetilde{z}^4},\qquad c=3/4,
\end{equation}
which trivially converges to the Wilkinson-Walker result in its regime $k\gg
1$, but better describes the results for all $k$. The results for the
regular region $\chi=-0.661$, $\eta\in [0.5, 1.0]$, are given by the crosses.
Once again, there is no universal result for the regular case.
Further, as expected, all three analytic functions (42)-(44) agree in the
large $k$ limit.

\subsection{Diagonal Matrix Elements: $P_{\widetilde{z}}(U_{nn}(z))$}

The previous results have been averages over various matrix elements. We now
show that the actual distributions of matrix elements can also be predicted by
universal functions. Consider, for example, the distribution of the matrix
elements $U_{nn}(z)=\langle\Psi_n(z)\mid\Psi_n(0)\rangle$. These can be seen to
be described by  a universal function for each $z$. The distribution
$P(U_{nn}(\widetilde{z}))$ is shown in Fig. 10 for $J^\pi=10^+$ states at
several values of $\widetilde{z}$. At $z=0$, the distribution is a
$\delta-$function centered at 1. As the separation of the wavefunctions
increases, the centroid shifts from 1 to 0,  and the overlap evolves to an
asymptotic  Porter-Thomas distribution (an approximate gaussian) at large $z$:

\begin{equation}
 P_{\widetilde{z}}(U_{nn}(z)) \rightarrow \left\{
   \begin{array}{ll}
       \delta(U_{nn}-1)       &  \widetilde{z} \rightarrow 0\\
       \exp{(-N U_{nn}^2/2)}  &  \widetilde{z} > 1 \\
   \end{array}\right.
\end{equation}
The large $z$ limit is not universal, as can be seen by the explicit
dimensional dependence of the result.  The Porter-Thomas result lies beyond the
range of universality which is largely $\widetilde{z}\leq 1$, but the precise
decorrelation of matrix elements within this range has the model independent
form shown in Fig. 10.

\subsection{Off-Diagonal Matrix Elements: $P_{\widetilde{z}}(U_{nm}(z))$}

A similar result exists for the distribution of off-diagonal matrix elements
$U_{nm}(z)=\langle\Psi_n(z)\mid\Psi_m(0)\rangle$ . The distribution
$P_{\widetilde{z}}(U_{nm}(\widetilde{z}))$ is shown in Fig. 11 for $J^\pi=10^+$
states at
several values of $\widetilde{z}$. At $z=0$, the distribution is also a
$\delta-$function, but now centered at $z=0$. As the separation of the
wavefunctions increases, the overlaps evolve  to an asymptotic  Porter-Thomas
distribution (an approximate gaussian) at large $z$:

\begin{equation}
 P_{\widetilde{z}}(U_{nm}(z)) \rightarrow \left\{
   \begin{array}{ll}
       \delta(U_{nm})       &  \widetilde{z} \rightarrow 0\\
       \exp{(-N U_{nm}^2/2)}  &  \widetilde{z} > 1 \\
   \end{array}\right.
\end{equation}
As before, the large $z$ limit is not universal, as can be seen by the explicit
dimensional dependence of the result. The solid curve in Fig. 11 is the
Porter-Thomas result.

\subsection{Correlations of Mean Fields}

Each value of $(\eta,\chi)$ corresponds to a deformed mean field characterized
by $(\beta,\gamma)$ determined from the minimum of Eq. (3). Because
wavefunctions decorrelate on order of $\widetilde{z}=\sqrt{D_2} z\sim 1$, the
actual correlation length in terms of the parameters $\eta$ and $\chi$
depends on spin, and is given by
$z\sim z_c\equiv 1/\sqrt{D_2(\Delta\beta,\Delta\gamma,J^\pi)}$. To explore the
spin dependence of the correlation length, we compute
$z_c$ in the IBM for $J^\pi=0^+,2^+,4^+,10^+$, and find typical values of
$z_c=0.16,0.14,0.11,0.05$. This has a roughly behavior
\begin{equation}
  z_c \sim 1 - \gamma J
\end{equation}
where $\gamma$ is a constant. How generic such an dependence
might be in other nuclear models is unclear, but it does indicate how rapid
states of different spin can decorrelate (n.b. the results can be corrected for
the  dependence of $z_c$ on $\sqrt{N}$, but this does not account entirely for
the  behavior). The equivalent values of $\Delta\beta$ and $\Delta\gamma$,
which correspond to statistically decorrelated configurations, depend rather
strongly on the parameter region. For example, in our calculations we can
obtain a range from $\Delta\beta=0.01$ to $1.3$, for the same correlation
length, depending on whether the shape is undergoing a rapid shape phase
transition or not in the particular parameter regime. One can only conclude
that near a shape phase transition, there can exist strong statistical
correlations between very distinct nuclear shapes.

\subsection{More Complicated Operators}

It is clear that one can explore many classes of operators and establish the
behavior of model-independent limits of those quantities. For instance in the
study of the $E2$ decay of high-spin states, Aberg\cite{aberg} has introduced
the matrix quantity
\begin{equation}
  T_{ij} = \mid\langle\Psi_i(J-2)\mid\Psi_j(J)\rangle\mid^2 (E_j(J)-E_i(J-2))^5
\end{equation}
where the parameter $J$ is the angular momentum, and is equivalent to $z$.
This matrix can be explored as a function of the correlation length, and has
different results in the chaotic limit, depending on the spin dependent
scaling $D_2$. One can consider other operators as well, and we would like to
point out that additional quantities can be constructed using our universal
predictions here, together with the analysis of Ref. \cite{walk},
which discusses how to compute arbitrary correlation functions.

\section{Multi-parameter Correlations: Topological Effects and Berry's Phase}

While formal studies of parametric correlations have been limited largely to
single parameter systems (see Ref. \cite{walk} for some exceptions), nuclear
deformation is usually described in terms of two or more shape parameters. When
two or more parameters are involved, one  finds that short distance
correlations can be modified by topological effects, due to Berry's phase. That
is, the correlation between quantities at $\beta,\gamma$ and $\beta',\gamma'$
depend on the path used to connect these points. Generally, for correlation
functions which are sensitive to phase information, we will show that
interference terms can strongly modify the expected results. We explore the
basic ideas here in the case of two parameters.

When a wavefunction undergoes parametric evolution on a closed circuit $C$,
it is well known that the wavefunction can pick up a topological phase:
\begin{equation}
  \Psi_n(z) \longrightarrow e^{i\gamma(C)}\Psi_n(z),
\end{equation}
where $C$ represents a loop in parameter space starting and ending at $z$.
For real symmetric matrices, such as our GOE ensemble, $\gamma(C)$ is only
$0$ or $\pi$ (mod $2\pi$)\cite{Berry}. Hence
\begin{equation}
\Psi_n(z+C) = \pm \Psi_n(z)
\end{equation}
where the sign depends upon the particular eigenstate and the path, and
$z+C$ represents the same point $z$ after following the closed loop $C$.
Of course, one does not have to follow a closed loop. A similar effect exists
if one follows
two distinct paths from $z$ to $z'$. Then phase differences result in
interference. Whether or not paths are in a chaotic or regular regime does
not change the flavor of the argument, but in the chaotic regime, more states
are likely to pick up a negative phase due to the many avoided level
crossings\cite{davida}.

\subsection{A 2-Parameter Random Matrix Model}

The simplest formulation of a two parameter correlated random matrix ensemble
is
\begin{equation}
 H(X,Y) = H_1\cos X  + H_2\sin X + H_3 \cos Y + H_4\sin Y
\end{equation}
where the constant random matrices $H_i$ are uncorrelated:
$\overline{H_iH_j}=\delta_{ij}$. It follows that
\begin{equation}
 \overline{H(X,Y) H(X',Y')} = \cos (X-X') + \cos ( Y-Y').
\end{equation}
Generalizations to arbitrary dimensions have been discussed by
Wilkinson\cite{walk}. We can now consider parametric excursions in the $(X,Y)$
plane, specifically two paths which connect $(0,0)$ to $(\Delta X,\Delta Y)$,
one of shortest length, and the other a longer path enclosing an area $A$.
Because the wavefunctions acquire a Berry's phase around the closed loop, which
can be $\pm 1$ for the GOE case, the area $A$ enclosed can modify expected
short distance behaviors.

\subsection{Correlations in the $\beta-\gamma$ plane}

We can now explore some of the topological effects in our two parameter theory
$H(\eta,\chi)$.  Consider a rectangular loop $C$ in parameter space which
encloses an area $A$. In analogy to scaled parameter $\widetilde{z}$, we define
the scaled area of  the loop as $\widetilde{A}=C(0)A\sim \widetilde{z}^2$. Then
an area  $\widetilde{A}\sim 1$ is a loop whose sides are approximately the
decorrelation length of observables. Such a loop stays within the universal
regime for all values of the parameter. In Fig. 12, we plot the distribution of
matrix elements $P_z(U_{nn})$ (see Eq. (45)) for such a loop. Starting from the
top of the figure, we have $\widetilde{z}=0$, and the distribution is a delta
function. As $\widetilde{z}$ increases, the distribution spreads in accordance
with universal predictions (cf. Fig. 10). At the farthest point of the loop,
the distribution is given by the middle figure. As the trajectory returns to
the
initial point, approximately half the eigenfunctions develop a negative
topological phase, and at the final point, which is precisely the initial
point, the distribution is equally split. All of the results in Fig. 12 are
within the universal regime, but one can see that topological effects can
destroy the expected behavior discussed in Fig. 10. For smaller loops, the
effect is smaller. The approximate behavior is\cite{davida}
\begin{equation}
  f = \left\{\begin{array}{ll}
   \frac{\widetilde{A}}{2} & \widetilde{A} \leq 1\\
   \frac{1}{2} & \widetilde{A} > 1\\
\end{array}\right.
\end{equation}
Here $f$ is the fraction of the total states which split to $-1$, and
$\widetilde{A}$ is the enclosed, scaled area.  The
fraction increases linearly with the area. Because saturation occurs near
$\widetilde{A}=1=C(0)\Delta\chi\Delta\eta$, and $C(0)\propto N$, the
size of the loop needed to see the maximal effect decreases  like $1/N$:
$\Delta\chi\Delta\eta\sim 1/N$. Hence we see that universal predictions
can be modified in multi-parameter theories due to the topology of the
parameter space. It is not sufficient to give only the metric distance in
parameter space in order to provide all universal predictions. One must
also consider the path taken to get to that metric separation.

So there are several aspects here to consider. Berry's phase effects are
independent of the underlying chaos along the chosen path, but depend more upon
the nature of the parameter space enclosed by the path. One could imagine a
loop in parameter space which is entirely regular, but encloses a chaotic
regime. The expression for the fraction of total states $f$ above assumes one
is always in a chaotic regime, and the enclosed area is also chaotic.

\section{Conclusions}

In conclusion, we have explored the adiabatic behavior of collective nuclear
excitations, and found that under the appropriate scaling of the parameter,
correlation functions and distributions of matrix elements behave universally.
Hence, if we wish to implement random matrix theory to study a complex nuclear
situation, we specify  immediately a multitude of model independent  results
related to the wavefunctions and energies. The results here indicate that a new
universality exists in nuclei, related to the `deformation' of the nucleus,
which is quite robust. As the random matrix predictions are generic, they
should be present in  other classes of nuclear states, generated from the shell
model, or other models. While we have focused on matrix element distributions
and certain correlation functions, it is clear that the scaling provides a
general type of approach to compute arbitrary correlation functions. This also
establishes that the use of random matrix theory with covariances of the type
(16) is quite reasonable.

We would like to thank R. Casten, C. Lewenkopf and V. Zamfir for useful
discussions, and M.Wilkinson for providing an advance copy of Ref.
\cite{walk}. This work was supported by DOE grant DE-FG02-91ER40608.

\newpage

{\noindent \Large \bf Figures} \\

\noindent Figure 1.  Distribution of matrix elements for the IBM in the chaotic
region (boxes, $\eta=0,\chi=-0.7$), and in the regular region (crosses,
$\eta=0.85,\chi=-0.661$). Both distributions are compared to the distribution
$P(|H|)\propto 1\sqrt{|H|} \exp(-|H|/V)$, with values of $V=10$ and $V=0.3$,
respectively.

\noindent Figure 2. Short distance behavior of the measured IBM
autocorrelation function $\langle H(z)H(0)\rangle$ in the
chaotic region (solid) and in the regular region (dots). The linear behavior
at small $z$ suggests $\alpha=1$ in Eqs. (21) and (26), which is not consistent
with the observed parametric energies $E_n(z)$. Hence $F_{ibm}$ cannot be
used as physical input into the random matrix theory through Eq. (16).

\noindent Figure 3. Instantaneous eigenstates of the Hamiltonian (1) for the
parameter range $\eta=0$ and $\chi$ as shown. (Top) Original energies; (Middle)
unfolded eneries; (bottom) Brody parameter along this path, indicating a rather
chaotic regime, $\omega=1$ being the GOE limit.

\noindent Figure 4. Instantaneous eigenstates of the Hamiltonian (1) for two
largely regular regions. (Left) A path from rotational $(SU(3))$ to vibrational
$(U(5))$ spectra through the regular region proposed recently\cite{Niall}. The
energies have been scaled by 1/2. (Right) The transition from vibrational
($U(5)$) to $\gamma-$unstable $(O(6))$ choosing a path which is weakly chaotic.
The average Brody parameters are $\overline{\omega}=0.23$ and
$\overline{\omega}=0.25$, respectively.

\noindent Figure 5. Slope-slope autocorrelation function, $C(\widetilde{z})$,
for the  parametric energies $\widetilde{E_n}(\widetilde{z})$ in the chaotic
(symbols)  and regular (dots) regions. The solid lines are the random matrix
predictions using (37) and (38) with $N=300$ and $N=50$, respectively. The
$0^+$ states have the poorest statistics, due in part to the small dimension of
the space (N=65).

\noindent Figure 6.  Scaling of $D_2=C(0)$ with $N$, the dimension of the
space. The boson number was varied as $N_b=10,15,20,25$, resulting in
dimensions $N=16,56,121,211$ for $J^\pi = 10^+$ states. The anticipated scaling
behavior, given by Eqs. (25)-(26) with $\alpha=2$, is linear,  shown by the
solid line. There are deviations at small boson number since the chaos is not
as strong there, and the dimensions are small.

\noindent Figure 7. (a) Analytic level curvature distribution $P(k)$ (solid)
compared to results of a GOE simulation (histogram). (b) Comparison of the
analytic distribution to those for $2^+$ states in the IBM (solid histogram).
The dashed histogram corresponds to $2^+$ states in the regular regime, and
has been scaled vertically by 1/5.

\noindent Figure 8. Wavefunction decorrelation function
$P_n(\widetilde{z})=\overline{|\langle\Psi_n(\widetilde{z})|
\Psi_n(0)\rangle|^2}$
for selected states in the chaotic regimes indicated (symbols), and one for
$2^+$ states in a regular region (dashes). The solid curves are our universal
predictions\cite{caio}. As there are no universal predictions in the regular
regimes, the dashed curve is only representative.

\noindent Figure 9. Distributions of off-diagonal matrix elements
$P_k(U_{nm}(\widetilde{z}))$ (where $k=|n-m|$). The IBM results for chaotic
$J^\pi=10^+$ states (boxes) agree well with the random matrix predictions
(solid) using $N=50$, as well as with a simple analytic  function (dashes). In
addition, we show the asymptotic results $P'_k(\widetilde{z})$ (dots) and
$P''_k(\widetilde{z})$ (dot-dashed) of Ref. \cite{walk} which converge for low
$k$ to the exact results. For reference, a similar calculation in the regular
region for $10^+$ states is included (crosses).

\noindent Figure 10. Distributions of diagonal matrix elements
$P_{\widetilde{z}}(U_{nn}(\widetilde{z}))$ at several values of $\widetilde{z}$
for $10^+$ state (solid) and the random matrix  predictions (dashed). The
distribution shifts from a delta function centered at one at $\widetilde{z}=0$,
to an asymptotic, non-universal Porter-Thomas distribution for
$\widetilde{z}\gg 1$

\noindent Figure 11. Distributions of off-diagonal matrix elements
$P_{\widetilde{z}}(U_{nm}(\widetilde{z}))$ at several values of $\widetilde{z}$
for $10^+$ state (histogram). The distribution shifts from a delta function
centered at zero at $\widetilde{z}=0$, to an asymptotic, non-universal
Porter-Thomas distribution for $\widetilde{z}\gg 1$ (solid).

\noindent Figure 12. Effects of Berry's phase on universal distribution
functions. The parameter $\widetilde{z}$ undergoes motion on a closed loop,
starting and ending at $\widetilde{z}=0$. Starting at the top, the distribution
evolves and eventually bifurcates due to the presence of topological phases.
Both results $\widetilde{z}=0.5$ are within the universal regime, but the lower
figure shows that the path taken to get to the point of interest can be
important. At the bottom, the fraction of matrix elements $f$ that change to
$U_{nm}=-1$ depends on the area enclosed. The minimum area needed to achieve
the maximum fraction of 1/2 scales as $1/N$, and is hence rapidly realized  in
large dimensional systems.

\end{document}